\documentclass[prints]{aa}
\usepackage{psfig}
\usepackage{supertabular}

\begin{document}
\title{Disturbed isolated galaxies: indicators of a dark galaxy
population?}
\author{I.D.Karachentsev\inst{1}
\titlerunning{dark galaxies}
\and V.E.Karachentseva\inst{2}
\and W.K.Huchtmeier\inst{3}}
\institute{Special Astrophysical Observatory, Russian Academy of Sciences,
	 N.Arkhyz, KChR, 369167, Russia
\and Astronomical Observatory of Kiev University, Observatorna 3, 04053
       Kiev, Ukraine
\and Max-Planck-Institut fur Radioastronomy, Auf dem Hugel 69,
    D-53121 Bonn, Germany}
\abstract{
    We report the results of our search for disturbed (interacting)
objects among very isolated galaxies. The inspections of 1050 northern
isolated galaxies from KIG and 500 nearby, very isolated galaxies
situated in the Local Supercluster yielded five and  four
strongly disturbed galaxies, respectively. We suggest that
the existence of ``dark'' galaxies explains the observed signs
of interaction. This assumption leads to a cosmic abundance of
dark galaxies (with the typical masses for luminous galaxies) that is
less than $\sim$1/20 the population of visible galaxies.
}
\maketitle

\section{Introduction}

 The paradigm of Lambda-CDM cosmology assumes that apart from dark halos
with normal (luminous) galaxies in their centers, completely dark clumps
(sub-halos) should also exist with masses $\sim(10^8 - 10^{11}) M_{\sun}$
(van den Bosch et al. 2003, Tully 2005, Yang et al. 2005). The current
understanding of the mass distribution and spatial distribution
of such invisible objects so far remains rather uncertain. However,
the total number of dark sub-halos may be comparable to or even
exceed in the number of the usual, luminous galaxies by a factor tens. For
instance, within volume of the Local Group, cosmological models of
structure formation predict about 300 satellites with masses greater than
$\sim3\cdot10^8 M_{\sun}$ (Klypin et al. 1999). This number is significantly
higher than the three dozen satellites actually observed in the Milky
Way and Andromeda. A possible explanation for this discrepancy is the
physical processes inhibiting star formation especially in low mass
clumps, thus implying the existence of a large number of dark satellites.

  Apparently, the completely dark ``galaxies'' (sub-halos) may be detected
via gravitational-lensing effects (Trentham et al. 2001). Recently
Natarajan and Springel (2004) measured masses of such substructures
in five clusters of galaxies utilizing archival Hubble Space Telescope
data and applying galaxy-galaxy lensing techniques. They found that
the fraction of total cluster mass associated with individual (visible and
invisible) sub-halos of $10^{11} - 10^{12.5}M_{\sun}$ ranges from 10-20\%.
But in practice, this approach is not efficient enough to explore dark
objects in loose groups and in the general galaxy field, because the
positions of dark galaxies on the sky are unknown a priori. Here, we
use another approach mentioned by Trentham et al. (2001), which relies
on searching for signs of non-motivated distortion visible on images of
spatially isolated galaxies.

\section{Interacting and isolated galaxies}

  As is well known, galaxies in close encounters show a significant
signature of gravitational interaction in the form of a
distortion of their structure, the presence of tails and bridges,
or a common diffuse envelope. All these features have been
quantitatively explained based on numerous N-body simulations ever
since Toomre \& Toomre  (1972).  Signs of interaction are seen in more
than 50\% of those binary galaxies where the separation is comparable to
the sum of the diameters (Karachentsev, 1987). In systems with greater
separation, as in triple galaxies, the relative number of interacting galaxies
is about 1/4 -- 1/3 . In magnitude-limited galaxy catalogs, the
fraction of interacting objects is equal to (8$\pm1$)\% (Karachentsev 1987).

  When considering more and more scattered systems of galaxies and single
galaxies of the ``general field'', one can expect a fraction of the
interacting objects among them to be nearly zero. This would occur if
there were no other objects except the luminous cataloged galaxies.
However, if completely dark galaxies (halos) with masses of $10^8 -
 10^{11} M_{\sun}$ exist, the phenomena of interaction  will occur in the
case of  extremely isolated galaxies, too.  Hence, an asymptotic relative
number of peculiar shapes among the most spatially isolated galaxies
may be a sensitive tool for estimating the cosmic abundance of massive
dark galaxies.

\section{Searching for distorted isolated galaxies}

  To search for such ``strange'' cases of interaction where the second
interacting companion is invisible, we used the ``Catalog of Isolated
Galaxies'' (Karachentseva 1973 = KIG). This catalog contains 1050 galaxies
without ``significant'' neighbors.
According to our estimates, the catalog objects do not suffer
essential perturbations from neighboring galaxies over some Gyrs.
As it is a sample of galaxies with $m < 15\fm7$ and declination $> - 2\degr$,
the KIG catalog includes only 4\% of the CGCG galaxies (Zwicky et al.
1961-1968); i.e it is a collection of a rather rare kind of galaxies.

  All KIG galaxies were inspected by us on the Digital Sky Survey 
(the blue POSS-II, when accessible). In some cases, we also studied galaxy
images in the 2 micron survey (2MASS) to check for the possibility of
double-nuclei systems as recent merging remnants. The results of our
inspections are presented in the upper part of Table 1. Its columns
contain: (1,2) galaxy numbers in the known catalogs taken from the
NASA Extragalactic Database (NED), (3,4) coordinates for the J2000.0 epoch,
(5) heliocentric radial velocity, (6) morphological type, (7,8) apparent
magnitude and angular dimension from the NED, (9) integrated blue luminosity
assuming a Hubble parameter $H_0 = 72$ km s$^{-1}$ Mpc$^{-1}$, (10) comments
relative to nuclear activity.  Footnotes to the table list basic
signs of peculiarity seen in the structure of the galaxies.

  Obviously, there are different factors of observational selection
affected the detection of isolated galaxies with peculiar shapes in
a sample limited by flux but not by volume. Therefore, we undertook
a new search for isolated distorted galaxies in a sample limited by
radial velocities, $V_{LG} < 3200$ km s$^{-1}$. In this sample of
$\sim$7500 galaxies covering the volume of
the Local Supercluster, about 60\% of the galaxies
reside in groups of different populations (Makarov \& Karachentsev, 2000).
The remaining $N\sim3000$ galaxies are characterized by
different degrees of isolation with respect to their neighbors
(with known radial velocities). We selected  500 of the most
isolated galaxies and inspected their images on DSS. Only 4 galaxies
out of 500 show significant signs of interaction. They are
listed in the bottom of Table 1. One case here, UGC 4722, turns out to
be common with the KIG sample. The DSS images of all 8 peculiar single
galaxies found by us are shown in Fig.1. The field of view in each
case is 8$\arcmin$ by 6$\arcmin$.

\section{Discussion}

  Signs of interaction between galaxies are  best developed when
the objects have similar masses. If the number of completely dark
sub-halos with typical masses of  $\sim10^{8} - 10^{11} M_{\sun}$ in any
volume approximately corresponds to the number of usual luminous
galaxies, then one may expect about 8\% interacting galaxies among
the isolated ones (interaction with an invisible object).  This rough
estimate ignores, of course, properties of spatial distribution of dark
sub-haloes with respect to the usual galaxies (i.e. the biasing problem).
The observed relative number of KIG galaxies with clear features of
interaction, 5/1050 = 0.5\%, turns out to be at least
one order of magnitude lower than expected.
Moreover, the observed frequency of disturbed shapes among
very isolated galaxies in the Local Supercluster, 4/500 = 0.8\%,
is consistent with the previous estimate made for the catalog sample.
In view of these results, it is rather unlikely that
the number of massive dark sub-halos is
similar to or even exceeds the number of the usual visible galaxies.

 It should be stressed that apart from interaction with a dark galaxy,
the observed morphological irregularities of isolated galaxies may
different origins:
a) there could have been a merger, with the companion now merged and not
visible anymore,
b) interaction with a companion now far away,
c) there could be large gas accretion from cosmic filaments, and asymmetrical
accretion could lead to perturbed morphologies (and also starburst
and AGN fueling).

In this context, we note that four galaxies out of eight listed in Table 1
are distinguished by their active nuclei. This may be a hint that their
peculiar shape is caused by a recent merging event. Moreover, some distorted
shapes of dwarf Sm galaxies may be generated by an asymmetric
star-formation burst inside the galaxy. Taking this possibility into account,
we suggest that the true relative number of isolated galaxies probably
disturbed by dark galaxies does not exceed a value of 0.3\%.

  The data in Table 1 show that peculiar isolated galaxies are
distributed over the sky rather inhomogeneously. This may indicate
the existence of clumps (filaments, clusters) in the distribution of dark
galaxies. The first observational evidence of the presence of a dark cluster
was made in the literature (Jee et al., 2005a,b).

  As noted by Neil Trentham (personal communication), dark galaxies
probably have low masses and negligible dynamical effects on massive galaxies
like the ones in the KIG sample. They can have much more substantial
effects on the nearby low-mass galaxies seen, for example, in the Catalog
of Neighboring Galaxies (Karachentsev et al. 2004). This sample contains
197 quite isolated galaxies with a "tidal index" $\Theta < 0$. (A negative
$\Theta$ means that the Keplerian cyclic period of the galaxy with respect
to its main neighboring disturber exceeds the cosmic Hubble time.)
The majority of them ($\sim$ 90\%) are low mass dwarfs. Recently,
Pustilnik et al. (2005) found that an isolated nearby galaxy, DDO 68 = VV 542
with $M_B = -14.3$ mag and $\Theta$ = -1.6, seems to be a disturbed object
"with a long curved tail on the South and a ring-like  structure at
the Northern edge." If DDO68 is a single such object in the
Local Volume, it yields a fraction of disturbed isolated galaxies,
1/197 = 0.5\%, the same as for the KIG catalog. We inspected
DSS images of all 197 isolated galaxies in the Local Volume and found
some more examples of distorted objects: NGC 1313 = VV 436 (IRAS source),
NGC 2537 = VV 138 (IRAS source), UGC 8837 = DDO 185. But, their peculiar
structure could be also understood as the result of galaxy merging
or an asymmetric star-formation burst.

  Apparently, the cases of isolated galaxies with distorted shapes,
as presented above, need special studies of their structure and kinematics.
Detailed investigations of such objects may give important
information on the population of dark galaxies.

\acknowledgements
{The authors thank  Neil Trentham, Francoise Combes, and Brent Tully
for useful comments. This search has made use of the
NASA/IPAC Extragalactic Database (NED), the STScI
Digitized Sky Survey (DSS), and the Two Micron All--Sky Survey (2MASS).
This work was supported by DFG--RFBR grant 03--02--06010.}

{}

\begin{table*}
\caption{ List of isolated galaxies with disturbed structure}
\begin{tabular}{llcrlccrl} \hline
 KIG &   Name   & RA(J2000.0)DEC &    $V_h$  &   Type  &   $B$   &  $a\times b$  &  lg$L$   & Comments   \\
     &          &                &   km s$^{-1}$ &    &  mag  &  arcmin &  $L_{\sun}$ &            \\
\hline
 293 & UGC 4722 & 090023.5+253641&  +1794 & Sdm&  15.2 & 1.6x0.2 &  8.9   &            \\
 341 & F635-02  & 093005.4+135821&     -  & Sa &  15.6 & 1.5x0.2 &   -    &            \\
 349 & UGC 5101 & 093551.6+612111& +11809 & Sb &  15.1 & 0.8x0.4 & 10.3   & Sy 1.5     \\
 940 & UGC11871 & 220041.4+103309&  +7978 & Sa &  14.5 & 1.1x0.7 & 10.5   & Sy 1.9     \\
 946 & UGC11905 & 220554.5+203822&  +7522 & Sa &  14.6 & 1.5x0.9 & 10.4   & II Zw 163  \\
\hline
  -  & ESO539-7 & 001848.5-190031&  +3187 & Sm &  14.8 & 1.5x1.3 &  9.5   &            \\
  -  & NGC 244  & 004546.3-153549&  + 944 & S0 &  13.8 & 1.2x1.0 &  8.9   & Haro 10    \\
 -   & ESO545-5 & 022006.1-194503&  +2338 & Sdm&  13.8 & 2.5x0.8 &  9.7   &            \\
\hline
\multicolumn{9}{l}{NOTES:}\\
\multicolumn{2}{l}{UGC 4722, RFGC 1465.}&
\multicolumn{7}{l}{LSB tail of 2$\arcmin$ directed from N-part
     of the galaxy to N.}\\
\multicolumn{2}{l}{F635-02.}&
\multicolumn{7}{l}{Wide curved tail to NW; or a diffuse companion in contact.}\\
\multicolumn{2}{l}{UGC 5101.}&
\multicolumn{7}{l}{Jet across the galaxy, longer to the W. Ring on the NW.
	    Merger? But a single nucleus in 2MASS.}\\
\multicolumn{2}{l}{UGC 11871, VV 812.}&
\multicolumn{7}{l}{Wide bright loop to S with a projected star.
	   Diffuse extension to NW.}\\
\multicolumn{2}{l}{UGC 11905.}&
\multicolumn{7}{l}{Long tail to S slightly curved to W. Loop on the NW,
	     and the LSB horizontal feature above.}\\
\multicolumn{2}{l}{ESO 539-7, DDO 1, UGCA 5.}&
\multicolumn{7}{l}{Asymmetric knotted arm at E-side.
	   $V_h = +2060$ (RC3).}\\
\multicolumn{2}{l}{NGC 244, VV 728, UGCA 10.}&
\multicolumn{7}{l}{Asymmetric envelope at E-side with a knot.}\\
\multicolumn{2}{l}{ESO545-05.}&
\multicolumn{7}{l}{ LSB loop from the core towards NE, short tail on the W.
		   A galaxy 10$\arcmin$ to E has}\\
& & \multicolumn{7}{l}{$V_h$ = 7608 km s$^{-1}$.}\\

\end{tabular}
\end{table*}
\begin{figure*}
\includegraphics{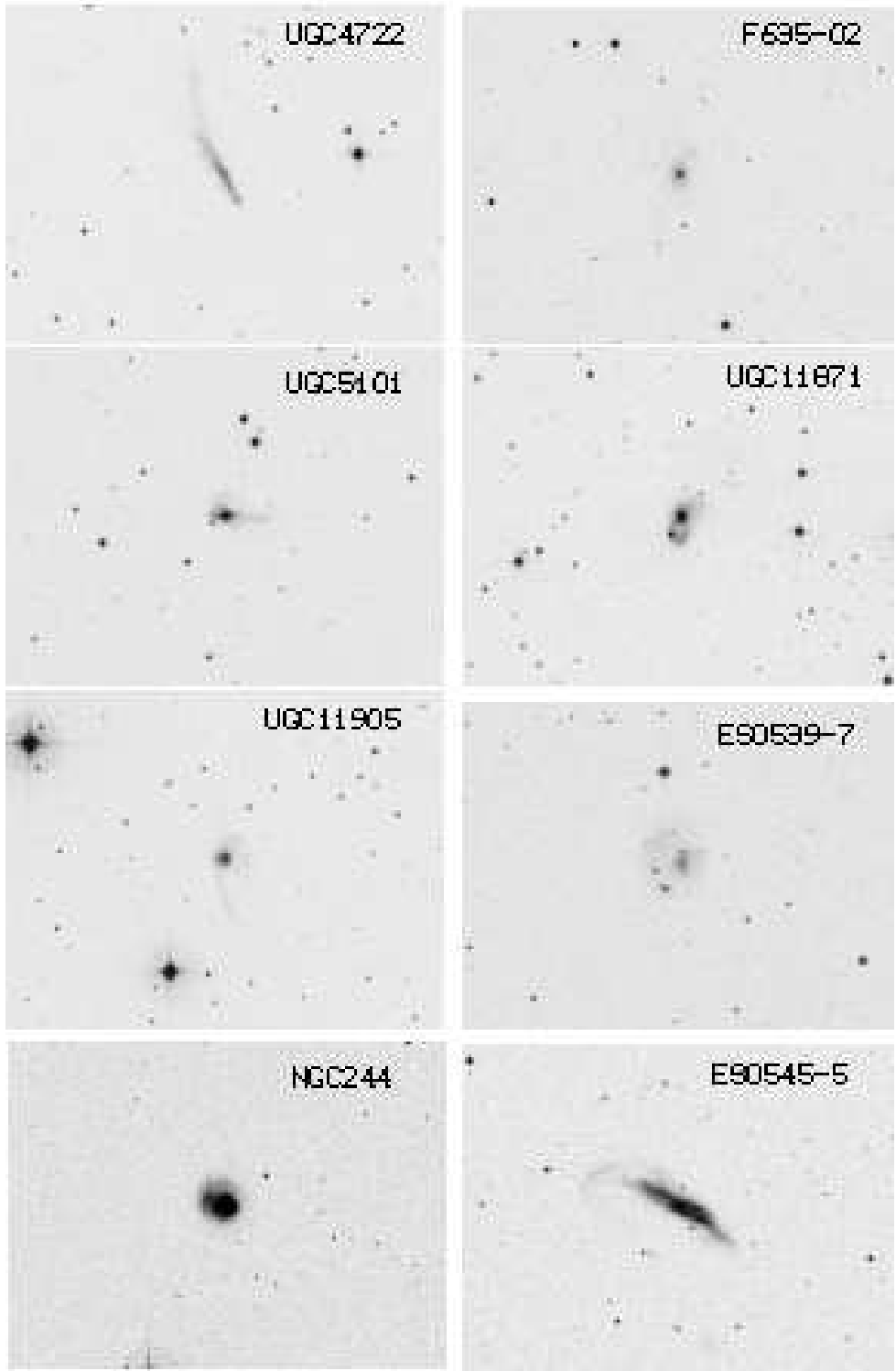}
\vspace{23cm}
\caption{STScI DSS optical images of the eight disturbed isolated
galaxies. East is left and North is up. The FoV is 8$\arcmin\times 6\arcmin$}.
\end{figure*}
\end{document}